# Brillouin scattering in multi-core optical fibers

Yosuke Mizuno[a)], Neisei Hayashi, and Kentaro Nakamura[a)]

*Precision and Intelligence Laboratory, Tokyo Institute of Technology,*

*4259 Nagatsuta-cho, Midori-ku, Yokohama 226-8503, Japan*

We measure the Brillouin gain spectra in two cores (the central core and one of the outer cores) of a ~3-m-long, silica-based, 7-core multi-core fiber (MCF) with incident light of 1.55 µm wavelength, and investigate the Brillouin frequency shift (BFS) and its dependence on strain and temperature. The BFSs of both the cores are ~10.92 GHz, and the strain- and temperature-dependence coefficients of the BFS in the central core are 484.8 MHz/% and 1.08 MHz/°C, respectively, whereas those in the outer core are 516.9 MHz/% and 1.03 MHz/°C. All of these values are not largely different from those in a silica single-mode fiber, which is expected because the cores are composed of the same material (silica). The difference in the BFS dependence of strain between the two cores may originate from the difference in structural deformation when strain is applied to the fiber, which is a unique characteristic to MCFs. The future prospect on distributed strain and temperature sensing based on Brillouin scattering in MCFs is finally presented.

---

[a)] Authors to whom correspondence should be addressed. Electronic mails: ymizuno@sonic.pi.titech.ac.jp and knakamur@sonic.pi.titech.ac.jp

With the emergence of next-generation internet technologies, such as online mobile applications and cloud computing, there is a growing demand for optical fibers with high transmission capacities that can slake our seemingly insatiable appetite for data. The spectral efficiency of these fibers has been extensively enhanced by using a variety of techniques such as coherent optical orthogonal frequency-division multiplexing, polarization-division multiplexing, and higher-order quadrature-amplitude modulation.[1] However, because of the nonlinear effects of optical fibers,[2–6] the information-transmission capacity of a standard silica-based single-mode fiber (SMF) is approaching its limit. A recent solution to this ever-increasing demand is space-division multiplexing based on multi-core fibers (MCFs).[7–15] By exploiting such multiple cores, researchers have been drastically enhancing the transmission capacity delivered over a single fiber.

In the meantime, Brillouin scattering in silica SMFs has attracted considerable interest[16-19] and has been applied to a number of devices and systems, including distributed strain and temperature sensors.[20-24] To improve the performance of these devices, Brillouin scattering properties in various special fibers have been investigated, and some of them have already been practically applied. Such special fibers include tellurite glass fibers,[25,26] chalcogenide glass fibers,[27,28] bismuth-oxide glass fibers,[26,29] photonic crystal fibers (PCFs),[30] rare-earth-doped glass fibers (REDFs),[31,32] polymethyl methacrylate (PMMA)-based polymer optical fibers (POFs),[33] and perfluorinated graded-index (PFGI) POFs.[34,35] Each special fiber has its own unique advantage; for instance, the Brillouin-scattered Stokes powers in tellurite and chalcogenide fibers are much higher than those in silica SMFs, owing to their large Brillouin gain coefficients[25–28]; the Stokes powers in REDFs can be adjusted by controlling the pumping light power (at 980 nm in erbium-doped fibers (EDFs))[32,37]; and Brillouin scattering in POFs is potentially applicable to high-sensitivity temperature measurement[35] as well as large-strain measurement.[36] Similarly, if Brillouin-scattered signals from multiple cores of an MCF can be simultaneously exploited, a discriminative measurement of strain and temperature[38–41] will be feasible by using a single fiber, as discussed below. To the best of our knowledge, the Brillouin properties in MCFs have not been experimentally investigated yet. Therefore, clarifying these properties is the first step to explore their potentials in practical applications.

In this letter, the Brillouin gain spectra (BGS) in two cores (the central core and one of the outer cores) of a ~3-m-long, silica-based, 7-core MCF are measured at 1.55 μm, and the Brillouin frequency shift (BFS) and its dependence on strain and temperature are investigated. The BFSs of both the cores are ~10.92 GHz, which are higher than that of a standard silica SMF by over 50 MHz. The strain- and temperature-dependence coefficients of the BFS in the central core are 484.8 MHz/% and 1.08 MHz/°C, respectively, whereas those in the outer core are 516.9 MHz/% and 1.03 MHz/°C. All of these coefficients are nearly identical to the values in a silica SMF, as is expected considering that the cores are composed of the same material, i.e., silica. The difference in the BFS dependence of strain between the two cores may be caused by the difference in structural deformation when strain is applied to the fiber. The future vision for MCF-based Brillouin sensors is then discussed.

Owing to an interaction with acoustic phonons, a light beam propagating through an optical fiber is backscattered, generating so-called Stokes light. This nonlinear process is known as Brillouin scattering,[16] where the central frequency of the Stokes light spectrum (called the BGS) becomes lower than that of the incident light. The amount of this frequency shift is referred to as the BFS, which is, in optical fibers, given as[16]

$$\text{BFS} = \frac{2\,n\,v_\text{A}}{\lambda}, \qquad (1)$$

where $n$ is the refractive index, $v_\text{A}$ is the acoustic velocity in the fiber core, and $\lambda$ is the wavelength of the incident light. As $n$ and $v_\text{A}$ are dependent on strain and temperature, the BFS also exhibits strain and temperature dependence, which serves as the operating principle of strain/temperature sensors. To date, the BFS dependence on strain and temperature has been investigated in a variety of special fibers. Table 1 summarizes the BFS and its dependence on strain and temperature, along with the refractive index, in silica SMFs,[17,18] tellurite fibers,[25,26] chalcogenide fibers,[27] bismuth-oxide fibers,[26,29] germanium-doped PCFs (main peak only),[30] REDFs (erbium-,[31] neodymium-,[32] and thulium-doped),[32] and PFGI-POFs.[35] Supposing that the wavelength of the incident light is 1.55 μm and that $n$ is not dependent on the wavelength, all the values in Table 1 have been calculated using Eq. (1). As can be seen, the BFS and its strain and temperature dependence drastically vary depending on the fiber materials and structures, which is extremely important in considering the applications of Brillouin scattering in MCFs.

The experimental setup for investigating the BFS dependence on strain and temperature in the MCF is depicted in Fig. 1, where self-heterodyne detection was used to observe the BGS with a high frequency resolution.[34] The output of a laser diode at 1.55 μm was divided into two, one of which was amplified with an erbium-doped fiber amplifier (EDFA) to 14 dBm and used as reference light. The other was also amplified with another EDFA to 30 dBm and injected into the MCF as pump light. The backscattered Brillouin Stokes light was then coupled with the reference light. The optical beat signal was then converted into an electrical signal with a photo detector (PD), and observed with an electrical spectrum analyzer (ESA) as BGS. For each BGS measurement, 20-times averaging was performed. The relative polarization state between the Stokes light and the reference light was adjusted with polarization controllers (PCs). The room temperature was ~28°C.

We employed a ~3-m-long, silica-based, 7-core MCF as a fiber under test,[11] the cross-sectional micrograph of which is shown in Fig. 2(a). The cladding diameter was 197.0 μm, and the core pitch was, on average, 56.0 μm. At 1.55 μm, the mode-field diameter, chromatic dispersion, and propagation loss of each core were 11.2 ± 0.1 μm, 18.5 ± 0.2 ps/nm/km, and 0.198 ± 0.011 dB/km, respectively. The cable cut-off wavelength of each core was 1.385 ± 0.026 µm. When the Stokes light returned from the central core (#1 in Fig. 2(a)) of the MCF was detected, as shown in Fig. 2(b), one end of the MCF was spliced to a ~1-m-long silica SMF (termed as SMF-1), which was connected to a circulator using an arc-fusion splicer (FSM-50S, Fujikura; central cores were automatically aligned), and the other end of the MCF was cut on an angle and immersed into matching oil ($n = 1.46$) to suppress the Fresnel reflection. In contrast, when the Stokes light returned from one of the outer cores (#2 in Fig. 2(a)) of the MCF was detected, as shown in Fig. 2(c), the outer core at one end of the MCF was spliced to one end of a ~1-m-long silica SMF (termed as SMF-2; manufactured by a company different from that of the SMF-1) using an arc-fusion splicer (S177A, Furukawa Electric; accurate core alignment was required). The other end of the SMF-2 was spliced to a ~1-m-long SMF-1. The Fresnel reflection at the other end of the MCF was suppressed in the same way. Strain was applied to the whole length of the MCF, two parts near the ends of which were fixed on translation stages by epoxy glue to avoid slipping. Temperature change was also applied to the whole length of the MCF, which was fixed on a heater.

First, the experimental results on the central core of the MCF are presented. The blue curve in Fig. 3(a) shows the BGS measured at room temperature. Two clear peaks (corresponding to BFSs) were observed at 10.87 and 10.92 GHz, which originate from the SMF-1 and the MCF, respectively. The BGS change when strain was applied only to the MCF is also shown in Fig. 3(a). With increasing strain, the BGS of the SMF-1 hardly changed, while the BGS of the MCF shifted to higher frequency. As the small peaks observed at 10.95 and 11.03 GHz did not exhibit strain dependence, they are the second- and third-order Brillouin peaks of the SMF-1, respectively.[16,19] The instability of the peak power of the MCF was caused by the spectral overlap with the higher-order peaks of the SMF-1 and by the polarization-dependent fluctuations. Using this result, we plotted the BFS of the central core of the MCF as a function of strain (Fig. 3(b)). The dependence was almost linear with a proportionality constant of 484.8 MHz/%, which is in agreement with the value (~493 MHz/%) in a standard silica SMF at 1.55 μm.[17]

We then measured the BGS dependence on temperature in the central core of the MCF (Fig. 4(a)). With increasing temperature, only the BGS of the MCF shifted to higher frequency. Figure 4(b) shows the BFS dependence on temperature, which was also linear with a proportionality constant of 1.08 MHz/°C. This value was almost the same as that (1.00 MHz/°C) in a standard silica SMF at 1.55 μm.[18]

Next, we present the measured results on the outer core of the MCF. The BGS measured at room temperature is shown in Fig. 5, where the Brillouin peak of the MCF was not clearly detected, because it overlapped with not only the peaks of the SMF-1 but also the peaks of the SMF-2 (including their higher-order peaks). To resolve this problem, we employed a differential measurement technique, with which the BGS of the SMF-1 and SMF-2 can be removed from the measured spectrum. The procedure was as follows: (1) obtain the BGS when local bending was applied to the SMF-2 at the region around the MCF/SMF-2 interface to induce considerable loss of over 40 dB; (2) obtain the BGS after the bending was released; and (3) subtract the BGS obtained in (1) from that obtained in (2). Note that the region to which bending was applied was so short (< 1 cm) that its influence on the measured results was negligible.

The BGS measured at room temperature using the differential measurement technique is shown as the blue curve in Fig. 6(a). The BGS of the MCF was detected separately from those of the SMF-1 and SMF-2. The BFS of the outer core of the MCF was ~10.92 GHz, which is almost the same as that of the central core. These values are higher than any other reported value in a standard silica SMF by > 50 MHz, indicating that the cores of this MCF have slightly higher degrees of hardness, and thus slightly higher acoustic velocities. This is most likely a result of the slight difference in material used to fabricate the MCF.

Subsequently, we measured the BGS dependence on strain in the outer core of the MCF, as also shown in Fig. 6(a). With increasing strain, the BGS of the MCF shifted to higher frequency. A dip independent of strain was generated at ~10.91 GHz by the differential measurement technique; along with the polarization-dependent fluctuations, this caused the instability of the peak power. The BFS dependence on strain in the outer core of the MCF is shown in Fig. 6(b). The proportionality constant was 516.9 MHz/%, which is ~30 MHz/% larger than that of the central core (484.8 MHz/%). This difference is not due to the measurement error (± ~5 MHz/%), which might be caused by the difference in structural deformation when strain was applied to the fiber.

Finally, the BGS dependence on temperature in the outer core of the MCF is shown in Fig. 7(a). With increasing temperature, the BGS of the MCF shifted to higher frequency. From the temperature dependence of the BFS (Fig. 7(b)), the proportionality constant was calculated to be 1.03 MHz/°C, which was almost the same as those in the central core (1.08 MHz/°C) and a standard silica SMF (1.00 MHz/°C) at 1.55 μm.[18]

In conclusion, we measured the BFS and its dependence on strain and temperature in the central core and one of the outer cores of a 7-core MCF. Although the strain dependence was clearly different between the central and outer cores, only slight difference was observed in the strain- and temperature-dependence coefficients of the BFS, which is expected because the MCF used in this measurement was fabricated with the intention that all the cores should be identical in material and structure. However, the difference in the Brillouin properties among the cores is expected to be enhanced by fabricating an MCF using different materials (see Table 1) and structures (note that several schemes have been developed to suppress the inter-core crosstalk by inducing phase mismatch among the cores).[12–15] If this difference is sufficiently large, by exploiting Brillouin scattering in multiple cores of the MCF, a discriminative measurement of strain and temperature distributions will be feasible only by use of a single fiber.[38–41] We anticipate that this paper will be an important archive exploring a new field of research: MCF Brillouin sensing.

The authors are indebted to Furukawa Electric Co., Ltd. for providing the MCF samples. This work was partially supported by Grants-in-Aid for Young Scientists (A) (no. 25709032) and for Challenging Exploratory Research (no. 26630180) from the Japan Society for the Promotion of Science (JSPS) and by research grants from the General Sekiyu Foundation, the Iwatani Naoji Foundation, and the SCAT Foundation. N. Hayashi acknowledges a Grant-in-Aid for JSPS Fellows (no. 25007652).

**Table**

TABLE 1. BFS at Room Temperature and its Strain and Temperature Coefficients in Silica SMF, Tellurite Fibers, Chalcogenide Fibers, Bismuth-Oxide Fibers, Germanium-Doped PCFs, Erbium-Doped Fibers, Neodymium-Doped Fibers, Thulium-Doped Fibers, and PFGI-POFs at 1.55 µm. The Refractive Index $n$ of Each Fiber is Also Presented.

| Fiber | BFS (GHz) | $n$ | Strain coefficient (MHz/%) | Temperature coefficient (MHz/K) |
|---|---|---|---|---|
| Silica SMF [a] | ~10.85 | ~1.46 | +580 | +1.18 |
| Tellurite [b] | ~7.95 | ~2.03 | −230 | −1.14 |
| Chalcogenide [c] | ~7.95 | ~2.81 | — | — |
| Bismuth-oxide [d] | ~8.83 | ~2.22 | — | −0.88 |
| Ge-doped PCF [e] | ~10.29 | ~1.46 | +409 | +0.82 |
| Er-doped [f] | ~11.42 | ~1.46 | +479 | +0.87 |
| Nd-doped [g] | ~10.82 | ~1.46 | +466 | +0.73 |
| Tm-doped [g] | ~10.90 | ~1.46 | +433 | +0.90 |
| PFGI-POF [h] | ~2.83 | ~1.35 | −122 | −4.09 |

[a]Refs [17,18], [b]Refs [25,26], [c]Ref [27], [d]Refs [26,29], [e]Ref [30], [f]Ref [31], [g]Ref [32], [h]Ref [35]

**Figure Captions**

Figure 1. Schematic of the experimental setup for Brillouin measurement.

Figure 2. (a) Cross-sectional micrograph of the 7-core MCF. Structures of the fiber under test for detecting Brillouin scattering in (a) the central core and (b) one of the outer cores of the MCF.

Figure 3. (a) BGS dependence on strain (0, 0.067, 0.135, 0.202, and 0.270 %) and (b) BFS dependence on strain in the central core of the MCF.

Figure 4. (a) BGS dependence on temperature (28, 40, 50, 60, 70, 80, 90°C) and (b) BFS dependence on temperature in the central core of the MCF.

Figure 5. Measured BGS in the outer core of the MCF, overlapped with those in the SMF-1 and SMF-2.

Figure 6. (a) BGS dependence on strain (0, 0.040, 0.069, 0.098, 0.126, 0.156, 0.184 %) and (b) BFS dependence on strain in the outer core of the MCF.

Figure 7. (a) BGS dependence on temperature (28, 40, 50, 60, 70, 80, 90°C) and (b) BFS dependence on temperature in the outer core of the MCF.

**Figures**

Figure 1.

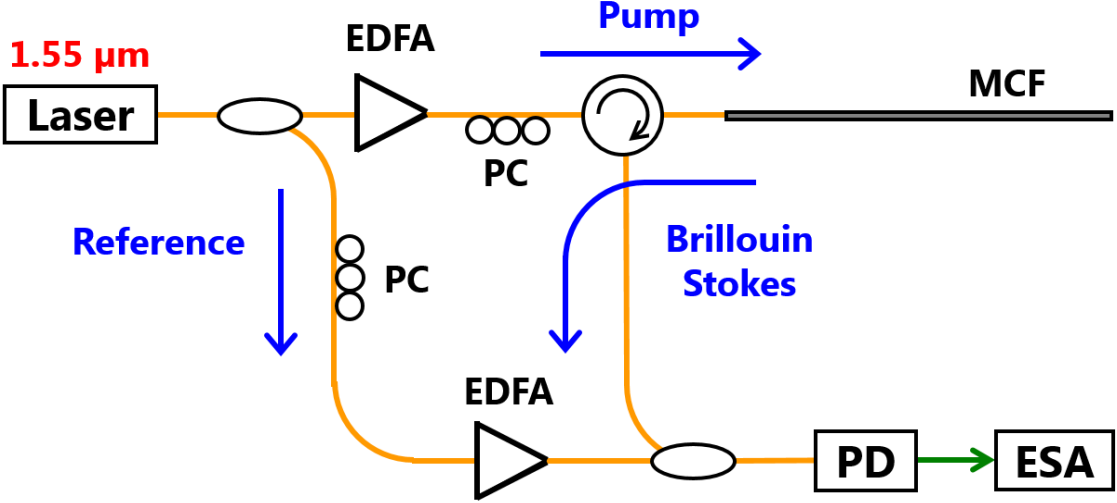

Figure 2.

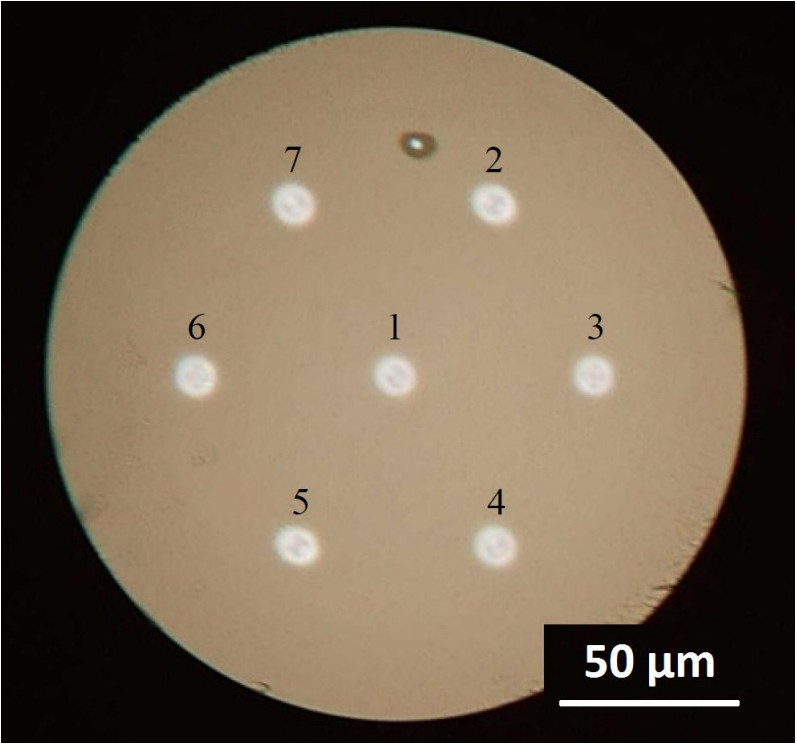

(a)

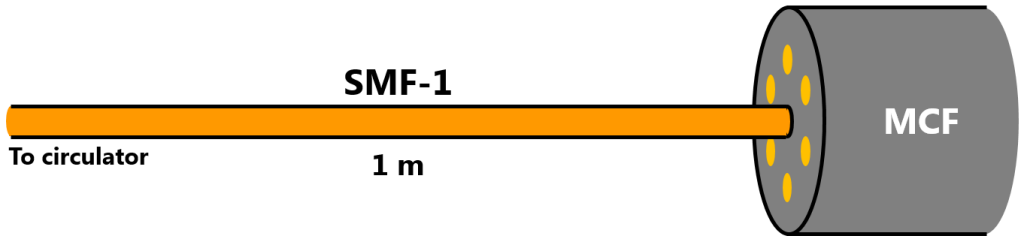

(b)

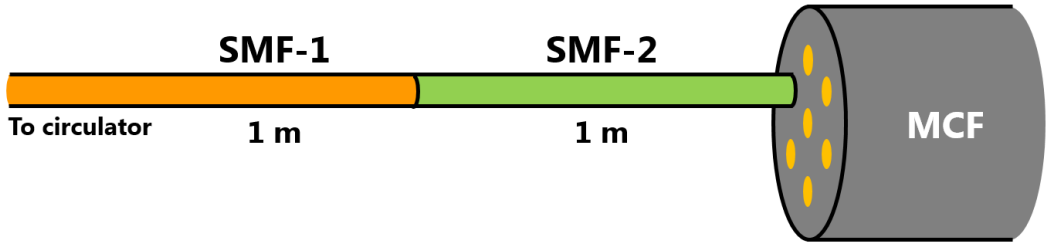

(c)

Figure 3.

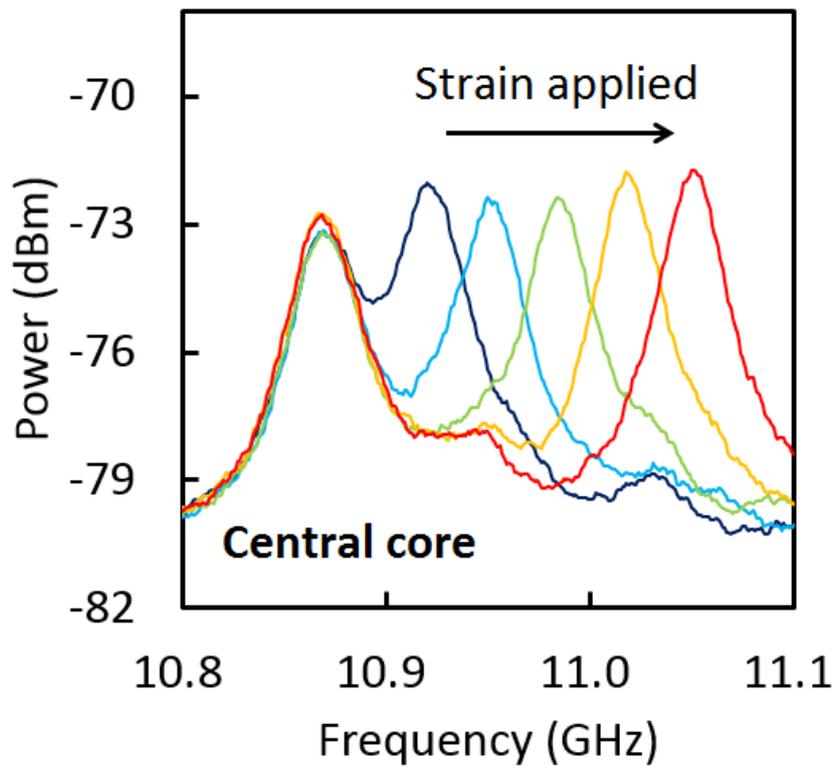

(a)

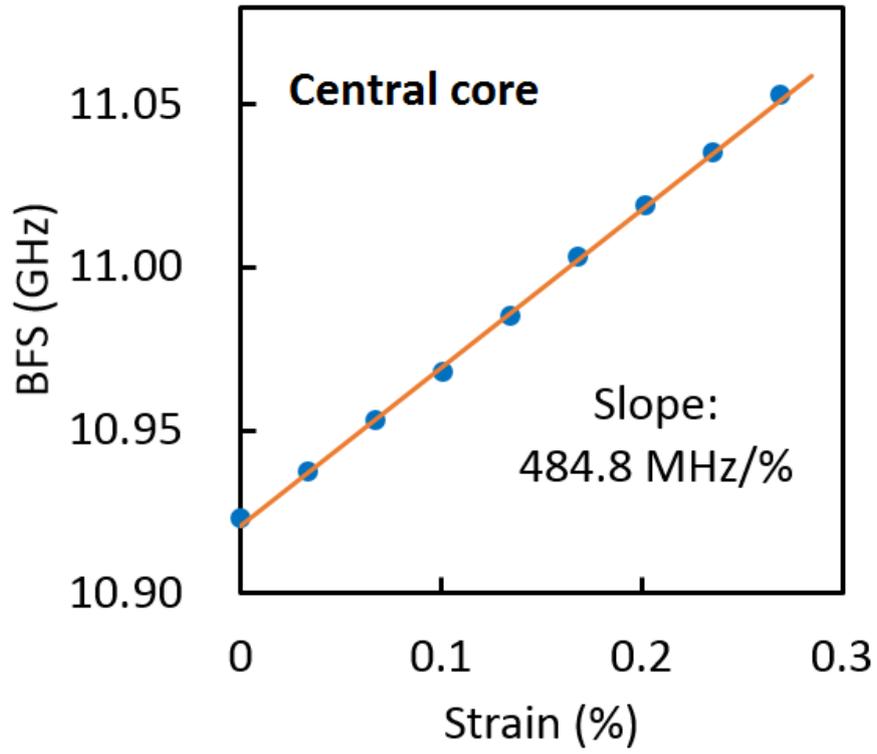

(b)

Figure 4.

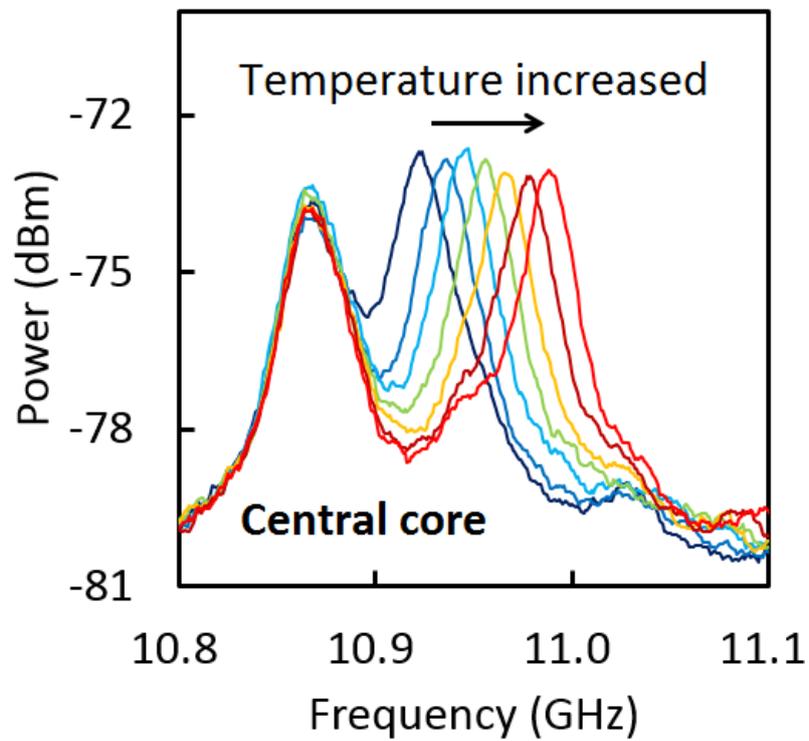

(a)

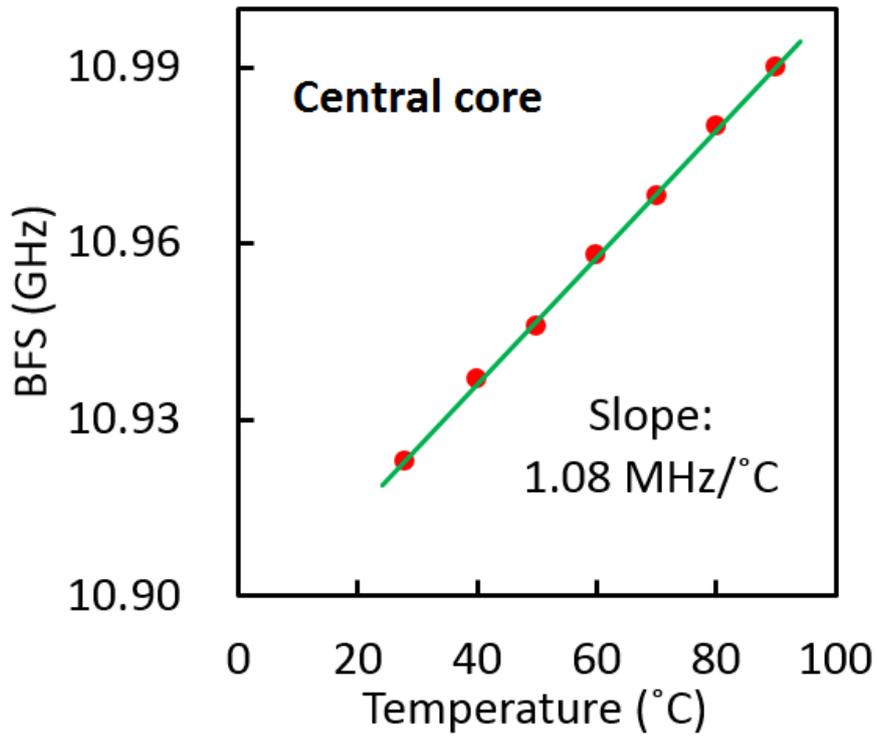

(b)

Figure 5.

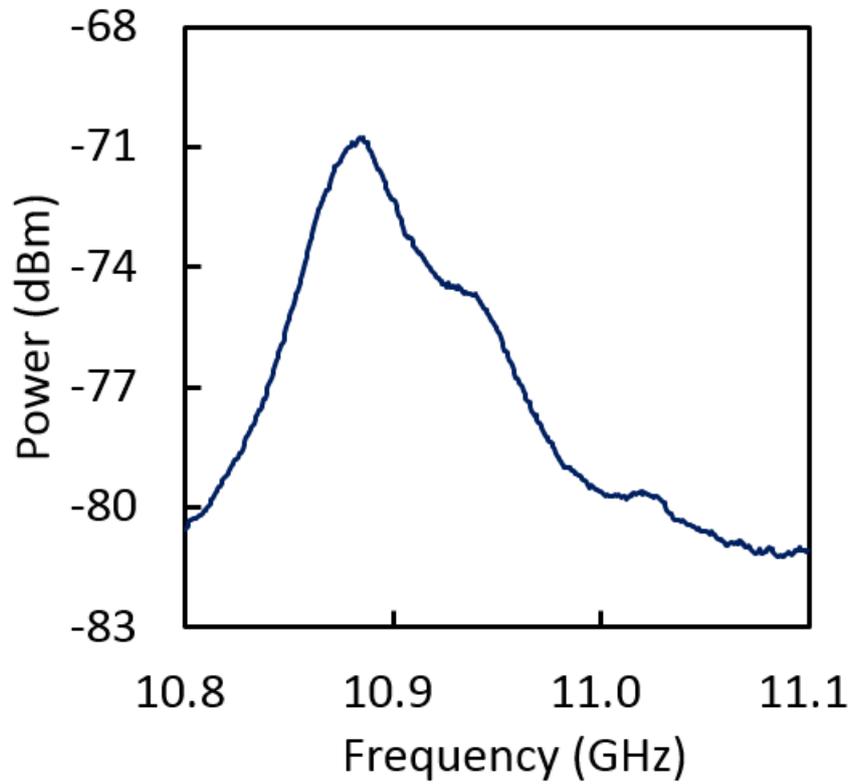

Figure 6.

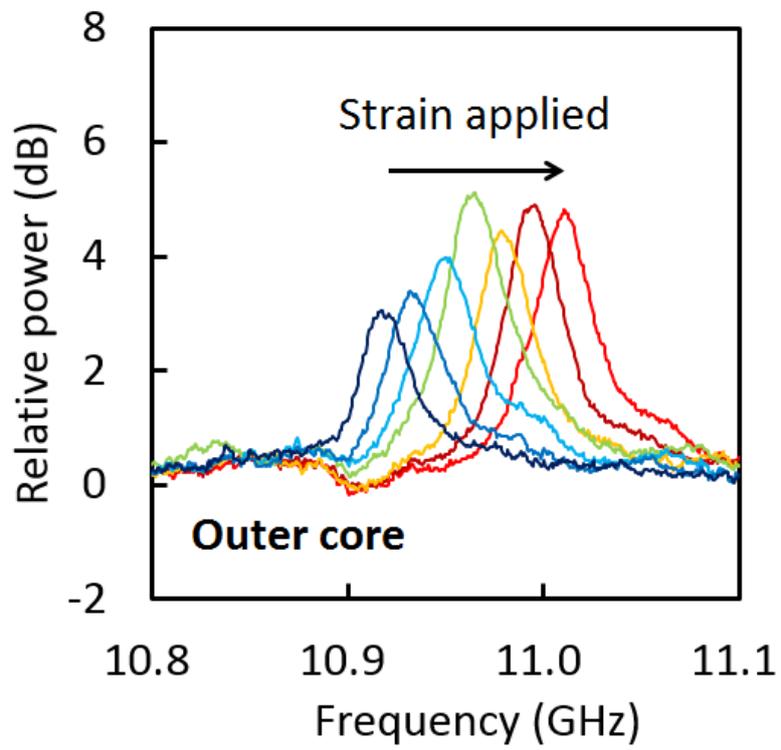

(a)

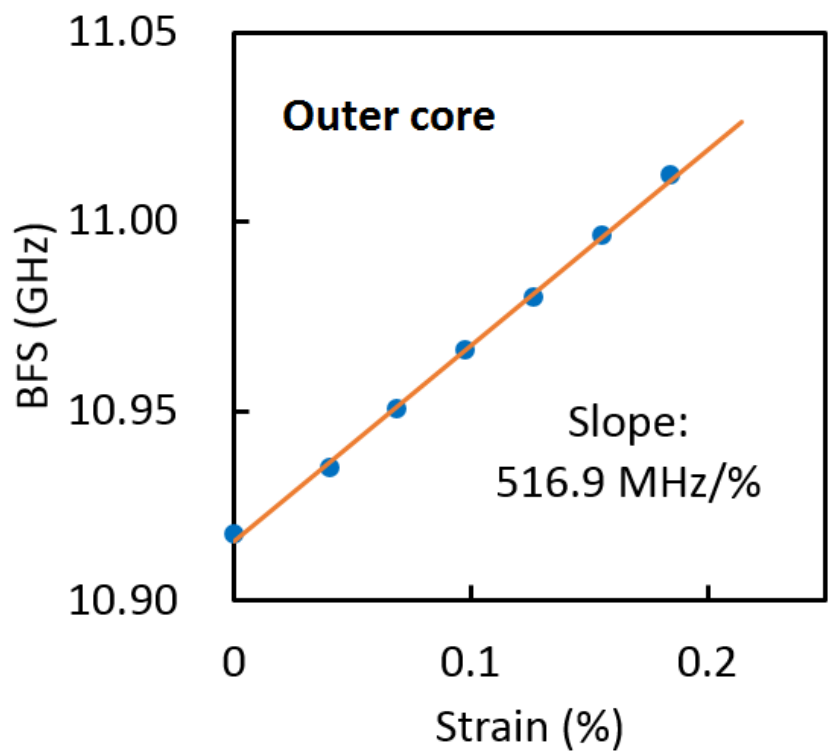

(b)

Figure 7.

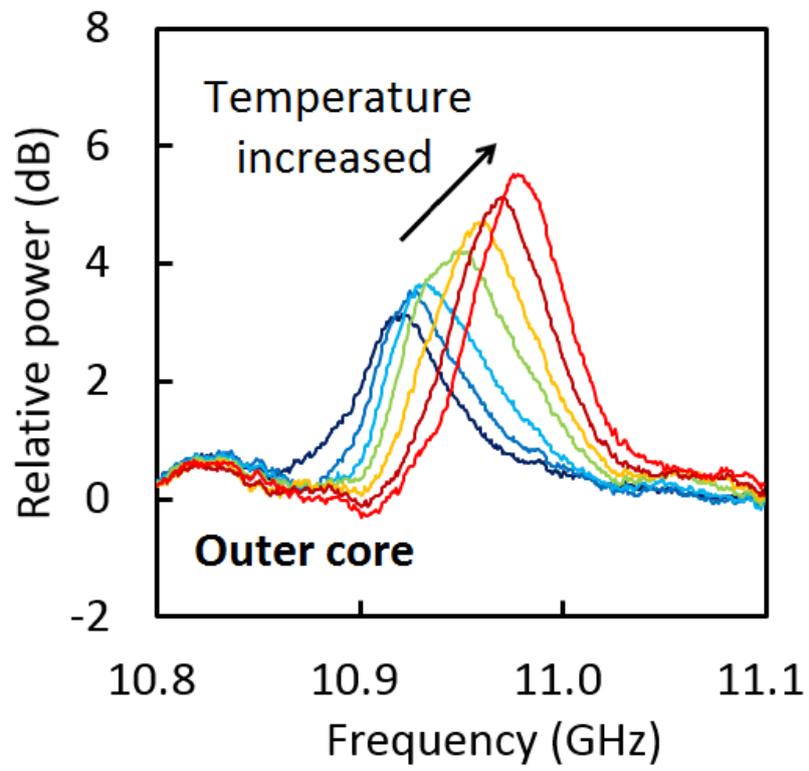

(a)

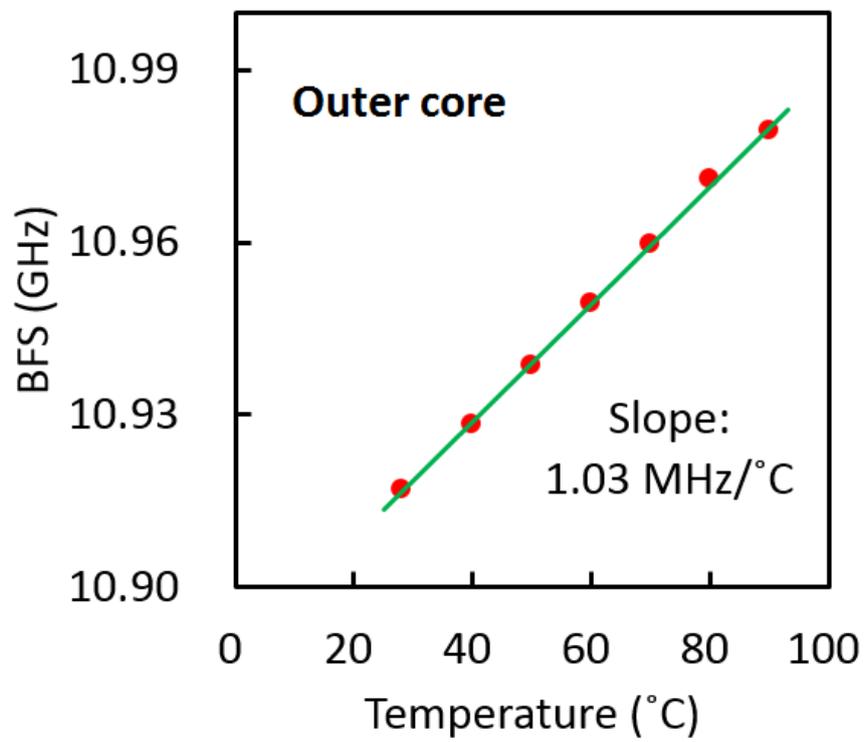

(b)